\documentclass[twocolumn,nofootinbib]{revtex4}
\newcommand\be{\begin{equation}}
\newcommand\ee{\end{equation}}
\newcommand\ba{\begin{eqnarray}}
\newcommand\ea{\end{eqnarray}}
\newcommand\eq{\begin{equation}}           
\newcommand\en{\end{equation}}

\usepackage{ulem} % \sout{this is wrong} to draw the line on top of the sentence
  
\input epsf
\usepackage{amssymb}
\usepackage{amsmath}
\usepackage{graphicx}

\begin{document}
\title{
%{\hfill  \small\\ ~\\~\\}
%Constraints on Light Magnetic Dipole Dark Matter from Linear Collider and Supernova
Constraints on Light Magnetic Dipole Dark Matter from the ILC and SN 1987A}
 \author{Kenji Kadota$^1$ and Joseph Silk$^{2,3,4}$\\
{ \small $^1$ \it   Department of Physics, Nagoya University, Nagoya 464-8602, Japan} \\
{ \small $^2$ \it Institut d'Astrophysique de Paris, CNRS, UPMC Univ Paris 06, } \\ 
{ \small \it UMR7095, 98 bis, boulevard Arago, F-75014, Paris, France} \\
 { \small $^3$  \it The Johns Hopkins University, Department of Physics and Astronomy, Baltimore, Maryland 21218, USA}\\
 { \small $^4$  \it Beecroft Institute of Particle Astrophysics and Cosmology, University of Oxford, Oxford OX1 3RH, UK}
}
%\date{\vspace{-5ex}}
%\date{}  % Toggle commenting to test
%\maketitle   
\begin{abstract}
To illustrate the complementarity of the linear collider and astrophysics bounds on the light (MeV-scale mass) dark matter (DM), we study the constraints on the magnetic dipole DM from the DM-electron interactions at the proposed International Linear Collider (ILC) and in supernova (SN) 1987A. We in particular focus on the $e^+ e^-$ annihilation which is the common process for producing DM pairs both at the ILC and in the SN. We estimate the bounds on the DM magnetic dipole moment from the mono-photon signals at the ILC and also from the energy loss rate due to the freely streaming DM produced in the SN. The SN bounds can be more stringent than those from the ILC by as much as a factor ${\cal O}(10^5)$ for a DM mass below $10^2$ MeV. For larger DM masses, on the other hand, SN rapidly loses its sensitivity and the collider constraints can complement the SN constraints.
\end{abstract}
\maketitle

\setcounter{footnote}{0} 
\setcounter{page}{1}
\setcounter{section}{0} \setcounter{subsection}{0}
\setcounter{subsubsection}{0}

\section{Introduction}
The nature of the dark matter (DM) remains an outstanding question which can provide crucial clues for the physics beyond the Standard Model (SM). In particular, besides the commonly studied weakly interacting massive particles (WIMPs) with weak scale mass, there has been growing interest in light DM whose parameter region has not yet been experimentally fully explored. For instance, current direct DM search experiments have recoil energy sensitivity down to of order a keV which limits the DM mass to be larger than about a GeV, and it would be of great interest to investigate the lighter mass range below a GeV for the potential window to new physics beyond the SM. Facing the wide open possibilities for the properties of DM, we study the interaction of MeV scale DM particles which possess a magnetic dipole moment and therefore interact with the photon. The DM magnetic dipole moment can be easily generated in many extensions of the SM such as asymmetric DM models and there have been many studies of the dipole DM, in particular for the light DM whose interactions with the SM particles can enjoy infrared enhancement due to the small momentum transfer in the photon exchange \cite{pos,kam,mas,kopp,dede,bank,zure2,ess,semi,light,barg2,heo3,paolo3,essi,ilidio,gary,nob}.

We aim to illustrate the complementarity of linear collider and astrophysical probes on light dark matter in the MeV-scale mass range. We focus on the interactions of the DM and electron/positron pairs and estimate how they can can affect the ILC and SN signals. For the ILC, we study the impact of the magnetic dipole DM on the mono-photon events where pairs of DM particles arise from $e^+ e^-$ annihilations, and, for the SN, we calculate the DM emission rate potentially affecting  SN cooling which is also due to  pairs of DM particles produced by $e^+ e^-$ annihilations. We relate the collider and SN phenomenology through this common DM production channel of $e^+ e^-$ annihilations and clarify how collider and SN signals can complement each other in constraining the DM magnetic dipole moment. 

\section{ILC and supernova constraints on the dark matter magnetic dipole moment}
We estimate the bounds on the magnetic dipole moment of MeV dark matter from $e^+ e^-$ annihilation which is the common channel for DM pair production both at the ILC and in supernova 1987A. 
\subsection{Magnetic dipole dark matter}
 We discuss the electromagnetic coupling for the interaction between DM and electrons, and consider fermionic DM $\chi$ whose gauge invariant coupling to the photon, up to dimension five, is via the magnetic dipole moment operator 
\ba
L=-\frac{i}{2} \mu \bar{\chi} \sigma_{\mu \nu} \chi F^{\mu \nu}
\label{lag}
\ea
$F^{\mu \nu}$ is the electromagnetic field tensor, and $ \mu$ corresponds to the magnetic dipole moment of DM and is parameterized as $\mu=1/\Lambda$ where $\Lambda$ represents the cut-off scale of the effective theory. $\Lambda$ for instance could be the mass scale of the charged particle running in the loop if this dimension-five dipole moment operator arises from the loop interactions with the heavy mediator. The ultraviolet-complete theory however is not sought in this paper in order to keep our model-independent discussions as general as possible. We also assume the only coupling between the DM and the SM particles is through the magnetic dipole interaction with photons given in Eq. (\ref{lag}) \footnote{The dipole moment in this paper refers to the magnetic dipole moment which preserves the discrete symmetries (C,P,T) and we leave the studies for the electric dipole moment which intrinsically breaks the P and T invariance for future work. Note a spin zero particle cannot have a permanent dipole moment either and the scalar DM coupling to the photon shows up only at dimension six 
%, $(\phi ^{\dagger} \overset{\text{\tiny$\leftrightarrow$}}{\partial_{\mu}} \phi)\partial _{\nu} F^{\mu \nu}$, 
which has been explored for instance in Ref \cite{foadi}. Another form of the electromagnetic interaction could occur if the DM is electrically charged, but the electric charge of the DM is severely constrained from the current experimental data and not pursued in our study \cite{mc}.}. The dipole interaction term vanishes for the Majorana fermion and we hence consider a Dirac fermion $\chi$ in this paper.
\subsection{Linear collider}
%The DM shows up as a missing energy at a collider and there have been studies on the DM search using the mono-jet evens with missing transverse energy at a hadron collider \cite{good10c,foxlhc,foxtev,ibe,belt,tim11,liam,paolokenji} and the mono-photon events with missing transverse energy at a lepton collider \cite{perel,fox1,chae,schm,bartel,dreilc,timcol,bart3}. 
The dark matter shows up as a missing energy at a collider and we consider the mono-photon signals $e^+ e^- \rightarrow \chi \bar{\chi} \gamma$ at the ILC %\cite{abe1,howie,beh,brau1,brau2,perel,fox1,schm,bartel,dreilc,timcol,bart3} to search for the dipole dark matter here.
\cite{brau1,brau2,bartel,abe1,fox1,timcol,bart3,dreilc,beh,howie,perel,schm} to search for the dipole DM.
The mono-photon, along with a dipole DM pair, at an $e^+ e^-$ collider can come from the initial state radiation (ISR) or from a final state DM via the dipole coupling. We study the mono-photon from the ISR which can dominate that from the final state DM due to the collinear enhancement when a photon is approximately collinear with an incoming beam. 
For the background, we simply consider the main irreducible background from the SM process $e^+ e^{-} \rightarrow \nu \bar{\nu} \gamma$.
To avoid the collinear and infrared divergences, we limit the phase space to be $E_{\gamma}>8$ GeV and $-0.995<\cos \theta_{\gamma}<0.995$. $E_{\gamma}$ has the $Z$ resonance peak around $\sqrt{s}/2 (1-M_Z^2/s)$ (242 GeV for the center of mass energy $\sqrt{s}=500$ GeV and 496 GeV for $\sqrt{s}=1$ TeV), and we hence impose a further cut $E_{\gamma}\leq 220 GeV$ for $\sqrt{s}=500GeV$ and $E_{\gamma}\leq 450$ for $\sqrt{s}=$1TeV to suppress the s-channel on-shell $Z$ recoil contributions. The background due to the $t$- and $u$- channel $W$ exchange can be further reduced by positively polarized electron and negatively polarized positron beams because of the $V$-$A$ nature of $W$ coupling. We adopt the beam polarization $P(e^+,e^-)=(-30\%,+80\%)$ for our analysis \cite{beh,ilcpol}. 

We implemented the magnetic dipole operator in Madgraph/Madevent and numerically obtained the upper bound on the dipole moment by requiring that 95\% confidence level upper limit on the background is smaller than the 95\% confidence level lower limit on the signal plus background \cite{mad5,bro,bae,kao2}. 
%We also assumed the integrated luminosity of $ L =250/$fb for center-of-mass energy of $\sqrt{s}=500$ GeV and $L =500/$fb for $\sqrt{s}=1$ TeV at the proposed ILC. 
Ignoring the systematic errors for simplicity, the constraints on the dipole moment can be obtained by requiring 
\ba
N_{sig}+N_{bg}-1.64 \sqrt{N_{sig}+N_{bg}}>N_{bg} +1.64 \sqrt{N_{bg}}
\ea
%which leads to
%\ba
%N_{sig}>(1.64)^2 \left( 1+2 \frac{\sqrt{N_{bg}}}{1.64}\right)
%\ea
%where $\sigma_{sig,bg},N_{sig,bg}$ represent the effective cross section after the cuts and the total number of events for the signal and SM background events. 
where $N_{sig,bg}$ represent the number of events after the cuts for the signals  and SM backgrounds \footnote{We simply consider the dominant statistical uncertainties in our estimation, which suffices for our purpose of quantitative estimation of the allowed magnitude of the dipole moment. We refer the readers to, for instance, Ref. \cite{dreilc,beh,barte} for more detailed studies on the model dependent systematic errors along with the full detector simulation and the optimized particle reconstruction.}.

\subsection{Supernova}
To study the complementarity to the $e^+ e^-$ linear collider, we consider $e^+ e^- \rightarrow \chi \bar{\chi}$ in the SN. %In this paper we  focus on the DM interactions with electrons/positrons in the nascent neutron star and we leave the DM-nucleons interactions in the SN along with the hadron collider complementarity for  future work. 
A pair of magnetic dipole DM particles can be produced from the annihilation of the relativistic $e^+ e^-$ in the same way as  DM pair production at the linear collider through the s-channel photon exchange. We estimate the emission rate of the freely steaming DM because any significant additional energy loss from the SN core could affect the shape and duration of the neutrino pulse from SN 1987A \cite{hira,bion}.
%The duration and shape of the neutrino pulse from SN 1987A can be well described by the analytical SN model and any significant additional energy loss from the SN core could alter the predicted neutrino pulse shape \cite{hira,bion}. 
The new DM channel can indeed be severely constrained not to conflict with the SN 1987A data, and the emission rates for SN 1987A have been extensively studied for the DM candidates such as axions, neutralinos and the gravitons in extra dimension models \cite{turn,keith4,gri,bet,lau,cul,han,drei2,das,sn1,kin,od,iwa,tubb,tub2,bar39,kache,dent}.

%The DM with the masses of order the SN core temperature can be thermally produced from the electron-positron annihilation in a SN $e^+ e^- \rightarrow \bar{\chi}+\chi$, even though the production of the DM with the mass exceeding ${\cal O}(10^2)$ MeV would be Boltzmann suppressed. 
The produced DM can escape the SN core if their mean free path is larger than the core size (of order 10 km), which can enhance the SN cooling rate and shorten the SN neutrino signals. For $e^{-} (p_1)e^+ (p_2)\rightarrow \chi (p_3) \bar{\chi}(p_4)$, the emissivity, the energy emitted per time and volume, is
\ba
\label{emi2}
\dot{\cal E}=\int d
\Pi_{i=1,4} \frac{d^3p_i}{2E_i (2\pi)^3}
(2\pi)^4 \delta^4  (p_1 +p_2- p_3- p_4)   \nonumber \\
f_1 f_2 (1-f_3)(1-f_4) |M|^2  (E_3+E_4)
\ea
where $f_i=[e^{(E_i-\mu_i)/T}+1]^{-1}$ is the Fermi-Dirac distribution function. 
%For the magnetic dipole coupling $-\frac{i}{2} \bar{\chi} \sigma_{\mu \nu} \mu_{\chi} \chi F^{\mu \nu}$, 
The matrix element squared, summed over the initial and final state spins, for the magnetic dipole interaction becomes
\ba
|M|^2
=\frac{64 \mu^2 e^2}{q^4} (p_1 \cdot p_2)[ (p_3 \cdot p_2 )  (p_4 \cdot p_2 )
\\ \nonumber
 +  (p_3 \cdot p_1 )  (p_4 \cdot p_1 )+ m_{\chi}^2 (p_1 \cdot p_2)]
\ea
where $q=p_1+p_2$, $m_{\chi}$ is the DM mass. We for simplicity ignore the electron mass $m_e \ll \sqrt{s}  $ in our analysis ($s$ is the usual Mandelstam variable representing the center of mass energy squared).  Performing the phase space integration leads to %under the constraint $s \geq  4m_{\chi}^2$ for the kinematic reason,
\ba
\label{intphase}
\dot{\cal E}=   \frac{2 \alpha  \pi^2 \mu^2 }{3}  \int_{4m_{\chi}^2}^{\infty}ds \int_{\sqrt{s}}^{\infty} dE_+ \int _{-{\sqrt{E_+^2-s}}}^{\sqrt{E_+^2-s}}dE_- 
\\ \nonumber
 s E_+ f_1 f_2  \sqrt{1-\frac{4m_{\chi}^2}{s}}  \left[1+\frac{8 m^2_{\chi}}{s} \right]
\ea
\ba
f_1= \frac{1}{e^{(E_++E_--2\mu_e)/(2T)}+1}, f_2= \frac{1}{e^{(E_+-E_-+2\mu_e)/(2T)}+1}
\ea
where $\alpha$ is the fine structure constant and we assume the complete final-state phase space is essentially available and ignore the Pauli blocking factors of the DM for simplicity.
In deriving the above expression, we changed the variables $(E_1,E_2,\theta)$ to $(E_+=E_1+E_2,E_-=E_1-E_2,s=2m_e^2+2E_1E_2-2p_1p_2\cos\theta)$ ($\theta$ is the angle between the three momenta ${\bf p}_1,{\bf p}_2$, and $p=|{\bf p}|$).
%$E_+\geq \sqrt{s},|E_-|\leq \sqrt{1-\frac{4m_e^2}{s}}\sqrt{E_+^2-s},s \geq  4m_{\chi}^2$.
We numerically integrate Eq. (\ref{intphase}) to obtain the emissivity. To obtain  reliable upper bounds on the dipole moment from the SN energy loss rate, one would need to implement the additional energy loss channel given above into a SN simulation code for various DM mass values. For the purpose of our paper to illustrate the compatibility of the astrophysical (SN) and collider (ILC) constraints on the DM properties, we instead perform the analytical estimation by applying the conventional Raffelt criterion which requires the energy loss rate due to the new channel to be less than $10^{19}$ erg/g/s not to invalidate the SN 1987A neutrino signals \cite{rafbook,rafan,raf4}. %This criteria should be used for the typical SN core condition. 

\subsection{Results}
\begin{figure}[htbp]
 \begin{center}
  \includegraphics[width=85mm]{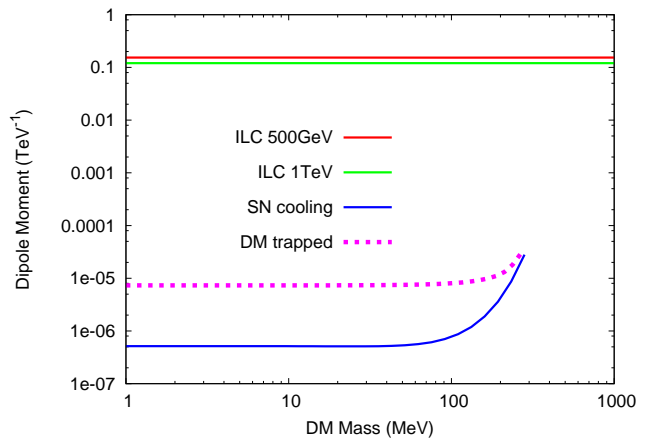}%{feb14.eps}{feb11b.eps}%{jan29.eps}%{jan27.eps}
 \end{center}
 \caption{The bounds on the DM magnetic dipole moment from the ILC and the SN. The regions above the ILC solid lines can be excluded from the mono-photon signal search. The SN energy loss due to the freely streaming DM excludes the region above the SN solid line for the excessive cooling and below the dotted line representing  DM trapping.}
 \label{fig1}
\end{figure}
The potential ILC sensitivity on the DM magnetic dipole moment is shown in Fig. \ref{fig1} for the polarized beams of electron and positron with the polarization $P(e^+,e^-)=(-30\%,+80\%)$ for $\sqrt{s}=$ 500 GeV and 1 TeV. The figure shows the bounds for $\sqrt{s}=500$ GeV with the integrated luminosity $250$/fb and $\sqrt{s}=1$ TeV with the integrated luminosity $500$/fb. The improvement on the dipole moment upper bounds is about 30\% by changing from (500 GeV, 250/fb) to (1 TeV, 500/fb). The sensitivity of the collider constraints on the DM mass is small, which is expected because $\sqrt{s}\gg m_{\chi}$. In fact, increasing the beam energy does not improve the constraints so much and the improvement mainly comes from the increase in the integrated luminosity.
%\subsection{trapping}
The energy loss per time per mass of the SN is $\dot{ {\cal E} }/\rho$ with $\rho$ being the mass density and the SN cooling constraints on the dipole moment from the Raffelt criterion $\dot{ {\cal E} }/\rho < 10^{19}$ erg/g/s is shown in Fig. \ref{fig1}. In our estimation, for concreteness, we use the constant density of $\rho = 3\times 10^{14}g/cm^3$, the core temperature $T=30$ MeV and the electron chemical potential $\mu_e=200$ MeV \footnote{There are ${\cal O}(1)$ factor uncertainties in the SN parameters which should be clarified from the detailed numerical simulations \cite{od,rafbook,fis}. We checked using $\mu_e =300$ MeV instead of $200$ MeV for instance increases the upper bound of the dipole moment by about a factor 2 for $m_{\chi}<100$ MeV, but our qualitative conclusion does not change due to those order unity uncertainties.}. The SN emission rate is not so sensitive to the DM mass when $m \ll T$, but its sensitively abruptly decreases once the DM mass exceeds ${\cal O}(10^2)$ MeV for a typical core temperature of $T\sim 30$ MeV for the kinematic reason. %This is expected because the relativistic electrons and positrons have the typical energy of order ${\cal O}(10-100)$ MeV around $T$, and the DM production process $e^+ e^- \rightarrow \chi \bar{\chi}$ suffers from the Boltzmann suppression once the DM mass exceeds around 100 MeV.
The SN bounds on the magnetic dipole moment turn out to be much tighter than those from the ILC for the lower mass range of $m_{\chi} \lesssim 10^2$ MeV by about a factor $10^5$. Because of the available energy range characterized by the typical core temperature adopted in our analysis $T=30$ MeV, however, the SN loses its sensitivity to a heavier dark matter mass exceeding a few hundred MeV where the ILC bounds can complement the SN bounds.

Before concluding our discussion, let us point out that the dipole constraints from the SN cooling in this paper are based on the energy emission rate due to freely streaming DM. For a large enough dipole moment, however, the DM diffuses instead of freely streaming, and we here give a simple estimation for the range for which our DM emission energy loss constraints are applicable. %The relevant interactions to estimate the DM mean free path involve charged particles (electrons/protons) because the dipole DM couples to the electromagnetic field. 
We estimate the elastic scattering cross section between the dipole DM and a charged particle through the photon exchange from ${d\sigma_{t}}/{d\Omega} \sim \alpha \mu^2 [m_{\chi}^4 +q^2 (-2m_{\chi}^2-m_t^2+s)-2m_{\chi}^2(m_t^2+s)+(m_t^2-s)^2] /(4 \pi s q^2)$ ($q$ is the 4-momentum transfer and $t$ represents the target particle (electrons and protons)) \footnote{In the calculation of this DM elastic scattering cross section, we ignored a number of complications such as the degeneracy effects, the form factors and coherence effects due to the small momentum transfer. More detailed studies covering those effects along with the thermal emission from the DM-sphere with a sufficiently large dipole coupling for the DM trapped regime will be presented elsewhere, which is model dependent requiring the specification of the non-constant density and temperature profile outside the inner core and the hadronic matrix elements.}. The mean free path $\lambda$ can be estimated as $\lambda= (\sum_t \sigma_{t} n_t)^{-1}$, and we assume the DM can be trapped in the core if the core radius ($\sim 10 $ km) is larger than $10 \lambda$. Then, by using the constant common number density $n_t \sim 10^{44}/m^3$ and a typical energy of the DM and electron ${\cal O}(10^2)$ MeV, we can estimate that the trapping condition is reached for the dipole moment of order ${\cal O}(10^{-5})$/TeV. For the parameter region $\mu \gtrsim {\cal O}(10^{-5})/$TeV, hence, our simple cooling arguments could fail and the DM diffusion would need to be taken into account. This limitation due to the DM trapping in the core is indicated with the dotted line in Fig. \ref{fig1} above which our SN cooling constraints could be obviated.\\
\\
In conclusion, we studied how the electron-DM interactions which show up for both the ILC and supernova can constrain the magnetic dipole dark matter to illustrate the complementarity of the collider and astrophysical probes on light DM in the  MeV mass range. We found the SN constraints turn out to be much tighter than the ILC ones for the lower mass range $m\lesssim {\cal O}(10^2)$ MeV, while the collider constraints can be complementary to the SN constraints for larger DM masses due to the rapid weakening of the SN cooling constraints for $m\gtrsim {\cal O}(10^2)$ MeV for a typical SN core temperature of $T \sim 30$ MeV.
Our study of MeV dipole DM deserves further investigation and we plan in the future to refine various uncertainties and simplifications made in our calculations. For instance, the detailed collider studies including the optimized selection cuts along with the detector simulations and particle identification/reconstruction efficiencies, such as those including the (detector specification dependent) background from $e^+ e^-\rightarrow e^+ e^- \gamma$ with both leptons missed, were left out in our estimates.
For the SN analysis, we paid particular attention to the DM-electron interactions in connection to the DM interactions at the ILC. We however note that DM-nucleon interactions could be important for the production and trapping of the DM in SN to tighten the bounds presented here, and we plan to perform more careful analysis for such issues in future work.
%These results can be compared with the previous literature which out the analogous constraints on the magnetic dipole moment from the other probes.  
%\section*{Acknowledgement}
\\
\\
This work was supported by the MEXT of Japan, the ERC project 267117 (DARK) hosted by Universit´e Pierre et Marie Curie - Paris 6 and at JHU by NSF grant OIA-1124403.
%%%%%%%%%%%%%%%%%%%%%%%%%%%%%%%%%%%%%%%

\end{document}